\documentclass[%
 aip,
 amsmath,amssymb,
 reprint,%
]{revtex4-1}

\usepackage{graphicx}
\usepackage{dcolumn}
\usepackage{bm}

\usepackage[utf8]{inputenc}
\usepackage[T1]{fontenc}
\usepackage{mathptmx}
\usepackage{etoolbox}
\usepackage{xcolor}
\usepackage{subcaption}
\usepackage{upgreek}
\usepackage{hyperref}
\hypersetup{
     colorlinks   = true,
     linkcolor    = blue,
     citecolor    = blue
}
\makeatletter
\def\@email#1#2{%
 \endgroup
 \patchcmd{\titleblock@produce}
  {\frontmatter@RRAPformat}
  {\frontmatter@RRAPformat{\produce@RRAP{*#1\href{mailto:#2}{#2}}}\frontmatter@RRAPformat}
  {}{}
}%
\makeatother
\begin{document}

\preprint{AIP/123-QED}

\title[]{Lithographically Defined Si$_3$N$_4$ Torsional Pendulum}
\author{Thomas Bsaibes}
\affiliation{Department of Physics, University of Maryland, College Park}
\affiliation{National Institute of Standards and Technology, Gaithersburg, MD}
\author{Charles Condos}%
\affiliation{Wyant College of Optical Sciences,University of Arizona}%
\author{Jack Manley}
\affiliation{National Institute of Standards and Technology, Gaithersburg, MD}%
\author{Jon Pratt}
\affiliation{National Institute of Standards and Technology, Gaithersburg, MD}
\author{Dalziel J. Wilson}
\affiliation{Wyant College of Optical Sciences,University of Arizona}%
\author{Jacob Taylor}
\affiliation{Department of Physics, University of Maryland, College Park}
\affiliation{National Institute of Standards and Technology, Gaithersburg, MD}
\affiliation{Joint Quantum Institute/Joint Center for Quantum Information and Computer Science, University of Maryland, College Park}%
\date{\today}

\begin{abstract} Torsion pendulums provide an opportunity to trap large masses in a potential weak enough to explore two-body gravitation. Cooled to, and then released from a ground state, weak quantum effects, including those from gravity, might reveal themselves in the evolving decoherence of a torsion pendulum, if its baseline dissipation were sufficiently dilute for quantum coherent oscillation. Monolithic ribbon-like, or multi-filar suspension geometries provide a key to such dilution in torsion, but are challenging to make. As a solution, we introduce a lithographically defined silicon nitride (Si$_3$N$_4$) ribbon suspension in a wafer-scale approach to pendulum fabrication that is conducive to such 2-D geometries, making extreme aspect ratios, and even multi-filar designs, a possibility. A monofilar, monolithic, centimeter scale torsion pendulum is fabricated and released in a first proof of concept. Mounted in vacuum, it is optically excited and cooled using measurement based feedback. Though only 37 mg,  the device displays a fundamental frequency of  162 mHz and an undiluted Q of 12$\, $000, demonstrating a foundational step towards ultra-coherent, ultra-low frequency torsion pendulums. 

\end{abstract}

\maketitle
\section{\label{sec:intro} Introduction}
Torsional pendulums have been used in experiments to measure Newton's gravitational constant \cite{quinn2014bipm,li2018measurements}, fifth force searches \cite{PhysRevD.86.102003,PhysRevLett.98.021101,shaw2022torsion,yang2025detecting}, and a variety of experiments probing the nature of gravity \cite{westphal2021measurement,10.1063/5.0145092,PhysRevD.111.082007}. They are well suited to measuring gravitational coupling, as is evident by equating the elastic potential of a traditional torsion balance, represented by its torsional spring constant $\kappa$, with the gravitational potential between a pair of spherical masses in near contact
\begin{equation}
  \kappa=G\left(\frac{\rho 4\pi}{3} \right)^2 r^5  
\end{equation}
where $r$ is the radius of the sphere, $\rho$ its density, and $G=6.674 \times 10^{-11}$ $\rm m^3 \rm s^{-2} kg^{-1}$ the universal gravitation constant. Assuming a hypothetical pendulum composed of tungsten spheres ($\rho= 2 \times 10^4$ $\rm kg\cdot \rm m^{-3}$) with a nominal radius of 10 mm,  a torsion pendulum has a stiffness equivalent to the two-body coupling when $\kappa \approx 10^{-11}$ $\rm N \rm m/ \rm rad$. Such potentials have been observed in previous pendulum experiments \cite{gillies1993torsion}.

For application to proposed tests of quantum gravity, an additional requirement is that the pendulum be capable of quantum coherent motion, which is likely to require a challenging combination of cryogenics and extraordinarily low dissipation \cite{Kafri_2014,nimmrichter2014optomechanical}.
To appreciate how low the dissipation must be, we define a quantum coherent pendulum as one that loses or gains less than one phonon to the environment per oscillation, typically expressed in the context of the standard quantum limit\cite{caves1980measurement} and at occupations well above one phonon as  
\begin{equation}
    Qf_\mathrm{n} \geq \frac{k_\mathrm{B} T}{h},
\end{equation}
where $f_\mathrm{n}$ is the natural frequency, $Q\equiv(2\delta)^{-1}$ a dissipation factor related to the loss angle $\delta$ that we typically interpret as the ringdown quality factor, $k_\mathrm{B}$ is the Boltzmann constant, $T$ is the temperature, and $h$ the Planck constant. The natural frequency is $f_{\rm{n}} =(2\pi)^{-1} (\kappa/I)^{1/2}$ where $I=8\pi \rho r^5/3$. Plugging in numbers for the hypothetical $r = 10$ mm tungsten masses yields a frequency of 123 $\upmu$Hz, or a period of 2.2 h.  Thus, a mechanical resonator suitable for quantum limited detection of gravitational coupling at a temperature of 10 mK must achieve a quality factor in excess of $10^{10}$, which has so far only been achieved in high frequency nanomechanical systems~\cite{dania2024ultrahigh,maccabe2020nano} with insufficient mass for studying gravitational interactions.



The performance of a torsional pendulum depends largely on the material and design of its suspension. High strength, and low intrinsic loss is desired; consequently many torsional pendulum based inertial sensors use tungsten \cite{PhysRevLett.98.021101,shaw2022torsion,10.1063/5.0145092,Guan:22} or fused silica fibers \cite{li2018measurements,yang2025detecting,westphal2021measurement,PhysRevD.111.082007,catano2020high}, where intrinsic loss is small, and $Q$ factors as high as $10^8$ have been reported in bulk samples\cite{ageev2004very}. The geometry of the suspension matters, with bifilar and multi-filar designs proposed in the past \cite{heyl1930redetermination,GGLuther_1999} to introduce a lossless gravitational restoring torque that supplants the elastic shear of the suspension. These unconventional torsion suspensions experience  gravitational dissipation dilution in a manner similar to that achieved in the pendulum suspensions of gravity wave detectors \cite{saulson1990thermal,huang1998dissipation}, but at sub-Hz frequencies without unwieldy size. The first and only successful demonstration of this strategy in a  torsion pendulum was a Cu-Be ribbon employed as the suspension in an apparatus to measure universal gravitation \cite{quinn2014bipm}. This device achieved a modest dilution factor of about 30. However, modern lithographic techniques have made it possible to make oscillators at the chip scale in a ribbon-like geometry \cite{PhysRevX.13.011018,pluchar2025quantum,engelsen2024ultrahigh,Shin:25,sementilli2022nanomechanical}, where a dilution factor as high as $3\times 10^4$ has been attained by exploiting residual stress\cite{hyatt2025ultrahigh}. In our system gravity would play the role of the residual stress providing the source of dissipation dilution. 
Lithographically defined suspensions allow for greater design flexibility, expanding the design space to include compact suspensions, extreme aspect ratios, and multi-filar suspensions.


This work reports the fabrication and characterization of a lithographically defined 25 mm long Si$_3$N$_4$ torsional pendulum; in our case the pendulum's bob is a rectangular torsion beam cut from a silicon wafer. The fabrication procedure yields a monolithic pendulum up to the final attachment point. The proof of principle pendulum exhibits a low torsional frequency of 162 mHz while maintaining a high intrinsic quality factor of $Q$ $\approx$ 12000. The success of this pendulum, and more generally this lithographic approach to torsion pendulum fabrication, hints that with realistic changes in ribbon geometry and mass loading, oscillators of exceptionally low frequency and high $Q$ should be possible, paving the way for experiments such as those proposed by Kafri and Taylor\cite{kafri2013noiseinequalityclassicalforces}, Kafri et al. \cite{Kafri_2014}, Carney et al.\cite{carney2021using}, and Kryhin and Sudhir \cite{PhysRevLett.134.061501}. In addition to the pendulum's torsional mode, we characterize two swing mode frequencies and demonstrate the ability to optically actuate the torsional mode oscillation of our device.

\section{\label{sec:fab} Fabrication}

Fabrication of the pendulum begins with a 400 $\upmu$m thick double-side-polished silicon (Si) wafer. Low pressure chemical vapor deposition (LPCVD) is used to deposit a 1.8 $\upmu$m thick layer of low stress (250 MPa) silicon nitride (Si$_{3}$N$_{4}$) on both sides of the wafer. A 2 $\upmu$m thick layer of silicon dioxide (SiO$_{2}$) is then deposited on the backside of the wafer (which defines only the frame and test mass) using plasma-enhanced chemical vapor deposition (PECVD) and serves as a hard mask to prevent the deep silicon etch from destroying the lithography pattern and the nitride.

The geometry of the device is laid out in a computer aided design (CAD) program, and the wafer is prepared for photolithography using a vacuum oven to apply a hexamethyldisilazane (HMDS) vapor to promote the adhesion of the photoresist to the nitride layer. The backside is then spin coated with a 5 $\upmu$m layer of high-viscosity positive photoresist (SPR220-7i) and a mask-less aligner is used to photolithographically transfer the CAD design to the photoresist layer on the wafer. The pattern is then developed and cured in a vacuum oven.

To transfer the lithography pattern to the substrate, the exposed oxide and nitride layers are removed in a reactive ion plasma etch. A protective layer of resist is then applied to the backside to protect the patterned nitride during topside processing. The photolithography step is then repeated using a layout similar to the topside that excludes only the ribbon; this layer defines the backside of the test mass and frame and serves as a mask for the potassium hydroxide (KOH) etch and constrains the thickness of the test mass to match our design parameters with the highest possible precision.

Once the backside pattern has cured, plasma etching is used to remove the nitride and oxide layers, thereby transferring the pattern to the silicon. We then remove 380 $\upmu$m of silicon using the Bosch procedure for deep silicon etching, cleave the wafer into individual chips, and strip any remaining polymer residue  with a commercial chemical specifically formulated for plasma etch residue removal. 
The chips are then prepared for KOH etching by cleaning them with solvents and blowing them dry with nitrogen. 

\begin{figure}
\includegraphics[scale = 0.45]{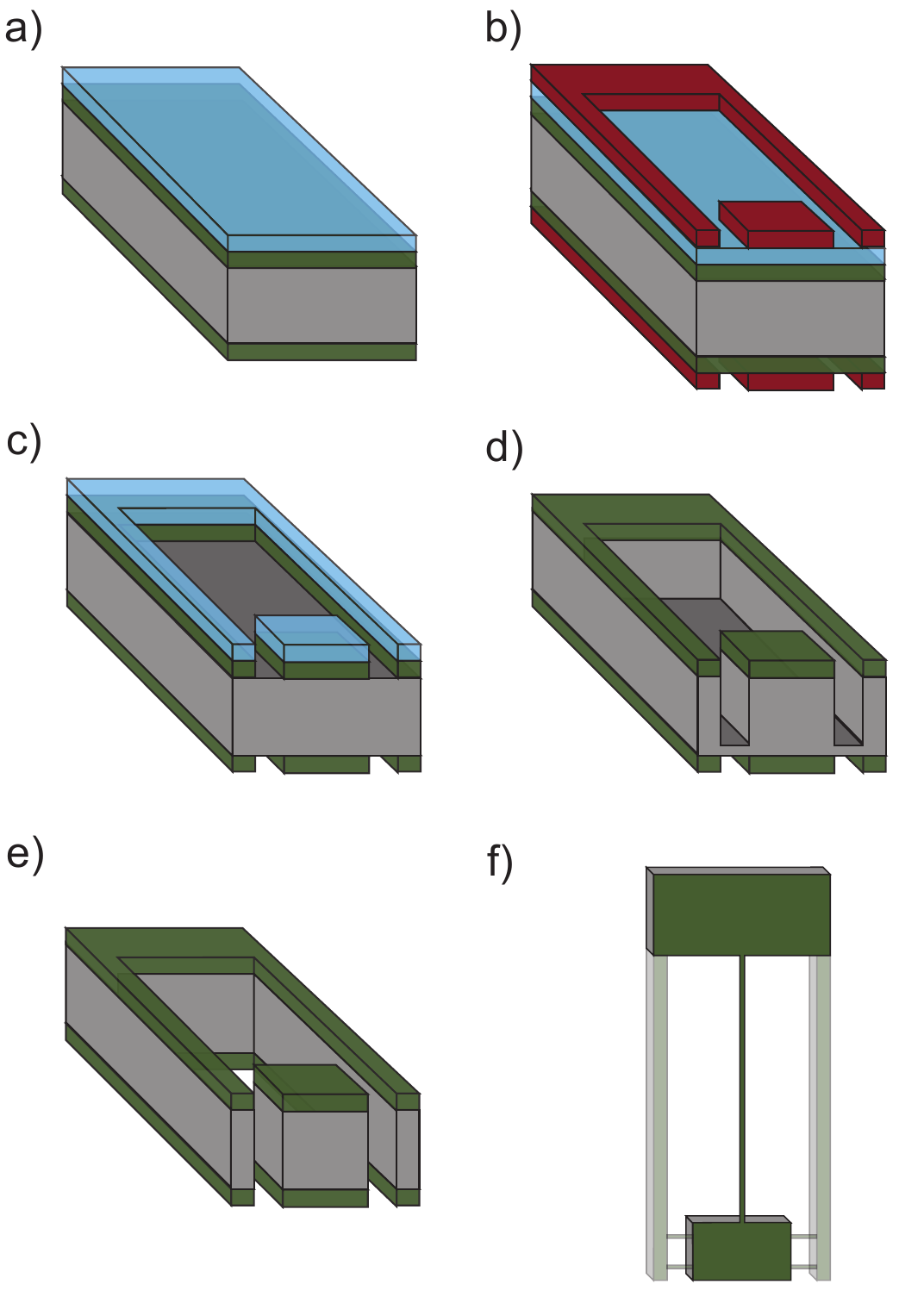}
\caption{Outline of the fabrication process: a) Si$_3$N$_4$ is deposited on both sides of a doubly polished Si wafer. b) A SiO$_2$ etch mask deposited on the backside of the wafer. c) Photoresist is spin coated and patterned using a mask-less aligner to define the suspension (topside only), frame, and test mass. d) The exposed areas are removed in a plasma etch. e) A deep silicon etch is used to remove most of the Si wafer. f) KOH is used to etch the remaining Si and release the pendulum; a final step, not shown, removes the breakout tabs and frame.}
\label{fig:fab process}
\end{figure}

To remove the remaining silicon and release the pendulum from its frame, we follow the KOH wet etching procedure developed at the University of Arizona outlined in Hyatt et al.\cite{jove68706}. Figure \ref{fig:fab process} outlines the fabrication process. 

\begin{figure}
\begin{subfigure}{0.5\textwidth}
\includegraphics[scale = 0.35]{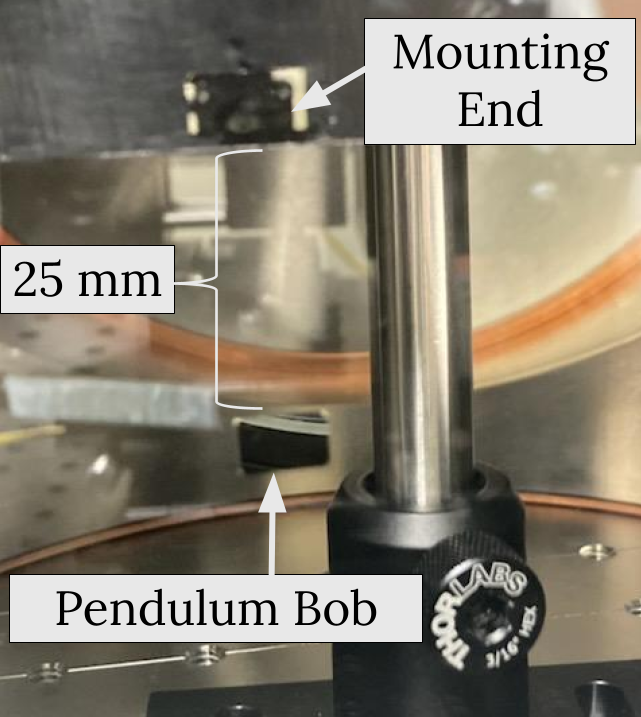}
\subcaption{}
\label{fig:unicorn_a}
\end{subfigure}
\hspace*{\fill}
\begin{subfigure}{0.5\textwidth}
\includegraphics[scale = 0.4]{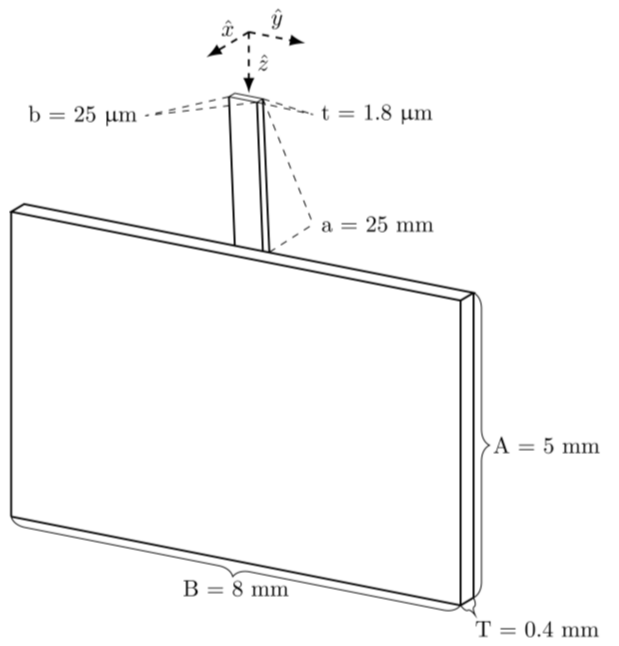}
\subcaption{}
\label{fig:unicorn_b}
\end{subfigure}
\caption{Photograph and schematic of the torsional pendulum. a) The bob of the torsional pendulum is an 8 mm $\times$ 5 mm $\times$0.4 mm Si wafer. At the top of the fiber is a piece of Si cut from the same wafer as the bob and used to mount the pendulum inside the vacuum chamber. b) A representation of the fiber and the bob of the torsional pendulum with their respective dimensions.}
\label{fig:unicorn}
\end{figure}

Following release, we use a diamond scribe to score and remove the breakout tabs and frame, which leaves a 25 $\upmu$m wide$\times$25 mm long$\times$1.8 $\upmu$m thick tether, with 8 mm $\times$5 mm$\times$ 0.4 mm rectangles at either end. One rectangle is used to mount the pendulum, while the other is the test mass (Fig \ref{fig:unicorn}). 

\section{\label{sec:model} Model}
We consider three modes of oscillation, two pendulum modes and a torsional mode. The torsional mode is a rotation about the $\hat{z}-$axis, see Fig. \ref{fig:unicorn_b}, while the pendulum modes correspond to $\hat{x}$ and $\hat{y}$ displacements. Due to the difference in the fiber's dimensions along the $\hat{x}$ and $\hat{y}$ directions, the stiffnesses in those directions are not the same and are referred to as the stiff pendulum mode and the soft pendulum mode for the swing about the $\hat{x}-$axis and the swing about the $\hat{y}-$axis respectively. The difference in stiffness and the moment of inertia lead to a 50 mHz frequency difference between the stiff and soft pendulum modes.

For calculating the stiffness of the pendulum modes, we follow the method outlined by Speake \cite{speake2018anelasticity}. Assuming the torque at the end of the fiber is negligible, a deflection described along either the $\hat{x}-$axis or the $\hat{y}-$axis is given by,
\begin{equation}
d_{i} = \frac{F_{i}}{\alpha_{i} W}\left\{\tanh(\alpha_{i} L)[\cosh(\alpha_{i} z)-1] + \alpha_{i} z - \sinh(\alpha_{i} z)\right\}
\label{eq.pend deflect}
\end{equation}
where $i\in\left\{\hat{x},\hat{y}\right\}$, $d_{i}$ is the deflection with the subscript indicating along which the direction the deflection occurs, $F_{i}$ is the deflection force applied at the end of the fiber, $W$ is the weight of the suspended mass, $L$ is the effective length of the pendulum, i.e the length of the fiber ($a$) plus half the height of the silicon bob ($A/2$), and $\alpha_{i} = \sqrt{W/(E_0H_{i}})$ with $E_0$ being the Young's modulus of Si$_3$N$_4$.  For low-stress low-pressure chemical vapor deposition (LPCVD) Si$_3$N$_4$ on Si, Kla{\ss} et al. infers $E_0 = 200$ GPa\cite{klass2022determining}; however, a wider set of values is often quoted, with 250 GPa being a common assumption for high stress LPCVD\cite{PhysRevX.13.011018} and the value used in this work. $H_{i}$ is the second moment of area and equals $bt^3/12$ for the soft pendulum mode and $tb^3/12$ for the stiff pendulum mode, where $b$ is the width of the fiber and $t$ its thickness, see Fig. \ref{fig:unicorn_b}. Taking the torque about the mounting point to be $\tau_{i} = F_{i}L$ and expressing the deflection of the end point as an angle through $d_{i} = \theta_{i} L$, the pendulum mode stiffness is worked out to be
\begin{equation}
k_{\text{pend}}= \frac{\alpha W L^2}{\alpha L - \tanh(\alpha L)}
\label{eq.kpend}
\end{equation}
where $k_{\text{pend}}$ is the stiffness of a pendulum mode.

The torsional stiffness is estimated using 
\begin{equation}
k_{\text{tors}} = \frac{JG_0}{a}
\label{eq.ktors}
\end{equation}
where $k_{\text{tors}}$ is the stiffness of the torsional mode, $G_0$ is the shear modulus of Si$_3$N$_4$ taken to be 125 GPa\cite{zerr2002elastic}, and $J$ is the torsional constant of a rectangle given as $J \approx \eta bt^3$. The constant $\eta$ is unitless and dependent on the ratio of the width to thickness ($b/t$); 0.312 is used as its value \cite{anwar2016structural}.

The predicted frequencies from the analytical model are shown in Table \ref{tab:model solutions}. We will see in section \ref{sec:meas} that the predictions match well with experiment.

\begin{table}
\caption{\label{tab:model solutions} Predicted stiffness, moment of inertia, frequency for three modes of oscillation.}
\begin{ruledtabular}
\begin{tabular}{llllr}
Mode &Stiffness\footnote{units: Nm/rad}&Moment of inertia\footnote{units: kg$\cdot$m$^2$} & Frequency\footnote{units: Hz} & \\
\hline
Pendulum Stiff& 1.053$\times 10^{-5} $ & 2.847$\times 10^{-8} $ & 3.061 & \\
Pendulum Soft & 1.008$\times 10^{-5} $ & 2.827$\times 10^{-8} $ & 3.005 & \\
Torsional     & 2.274$\times 10^{-10}$ & 2.008$\times 10^{-10}$ & 0.169 & \\
\end{tabular}
\end{ruledtabular}
\end{table}

\section{\label{sec:meas} Measurements}
\begin{figure}
\includegraphics[scale=0.4]{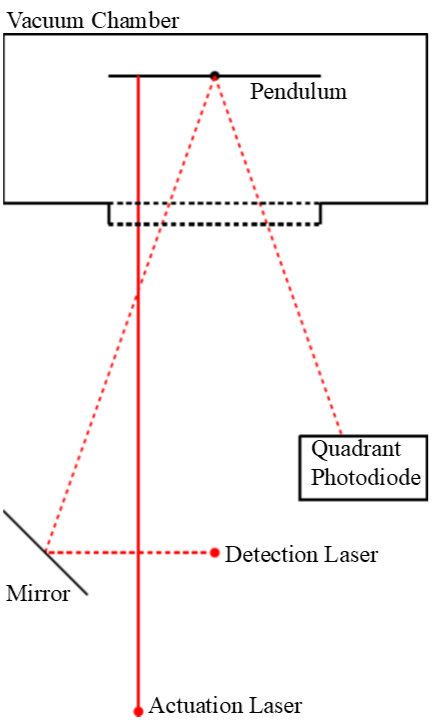}
\caption{Schematic of the experiment. The pendulum, in vacuum, is about 300 mm from the optical set up. The detection laser (red) is directed to the pendulum bob via a mirror and the bob's motion is recorded with a quadrant photodiode. 
}
\label{fig:schematic}
\end{figure}

The pendulum is mounted in a high vacuum chamber and its motion is monitored using an optical lever through a viewport. A schematic of the experiment is shown in Fig. \ref{fig:schematic}. The experiment makes use of two lasers: a detection laser and an actuation laser. The detection laser is directed to the pendulum's bob via a mirror, and the bob's motion is recorded using a quadrant photodiode; while the actuation laser is used to apply an optical force to the torsional mode of the pendulum.

\subsection{\label{subsec:noise} Noise Analysis and Ring-down}
\begin{figure}[t]
\includegraphics[scale = 0.17]{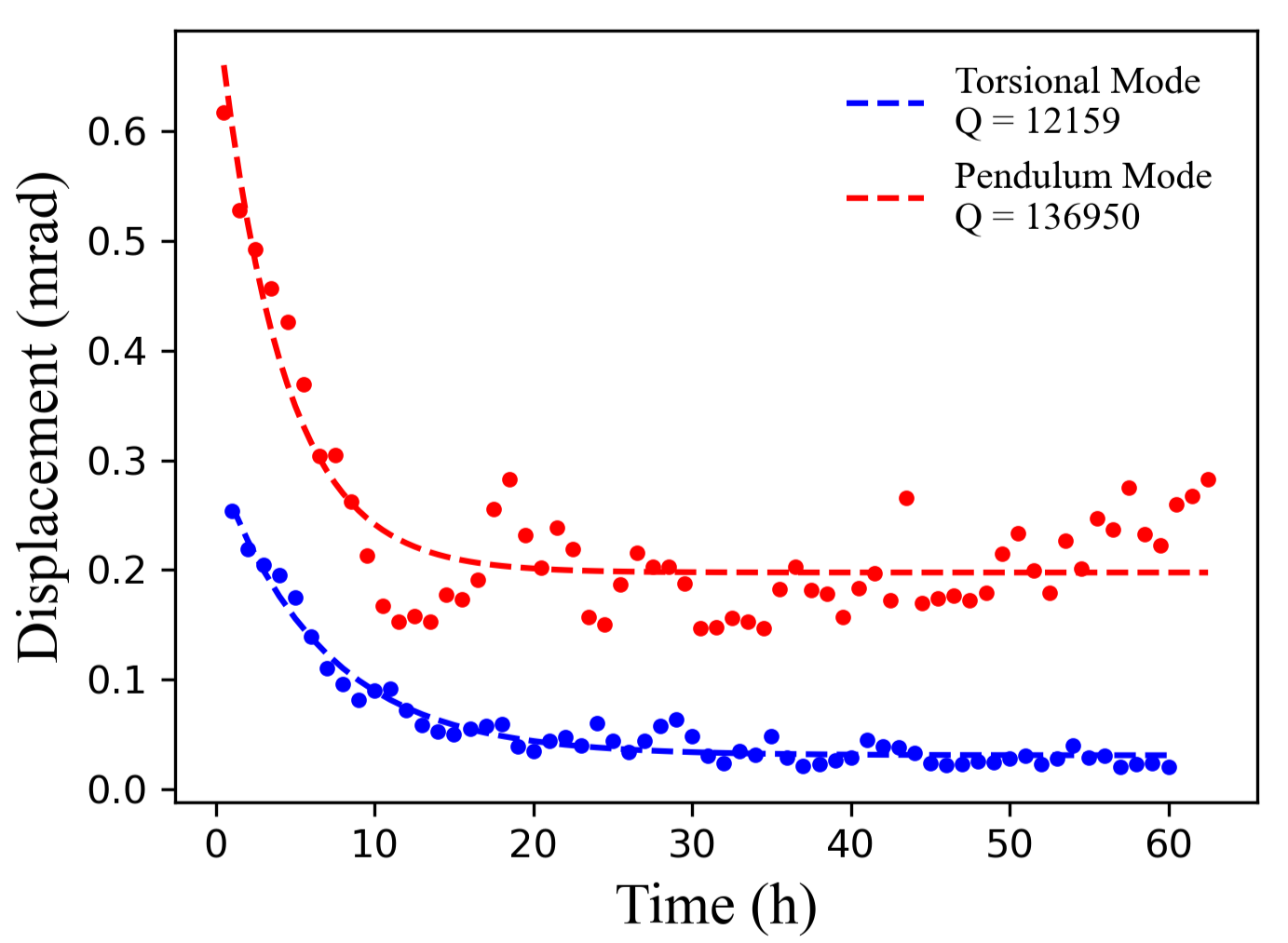}
\caption{Ring-down of the torsional and pendulum modes and their respective fits with extracted quality factors.}
\label{fig:ringdown}
\end{figure}

The pendulum was rung-up using optical actuation (section \ref{subsec:actuation}) and allowed to freely ring-down over 60 hours. The ring-down measurement was carried out stroboscopically, with the detection laser being turned on for 40 s every hour. The root mean square (RMS) amplitude of the signal was computed over 40 s and fit to an exponential decay. The damping rate, $\gamma$ is extracted from the fit. Figure \ref{fig:ringdown} is an example of a typical ring-down measurement for the pendulum mode and the torsional mode. They yield $Q$ values of $\approx$ 137$\, $000 and $\approx$ 12$\, $000 respectively.

\begin{figure*}
    \centering
    \begin{subfigure}[]{0.49\textwidth} 
        \centering
        \includegraphics[width=\textwidth]{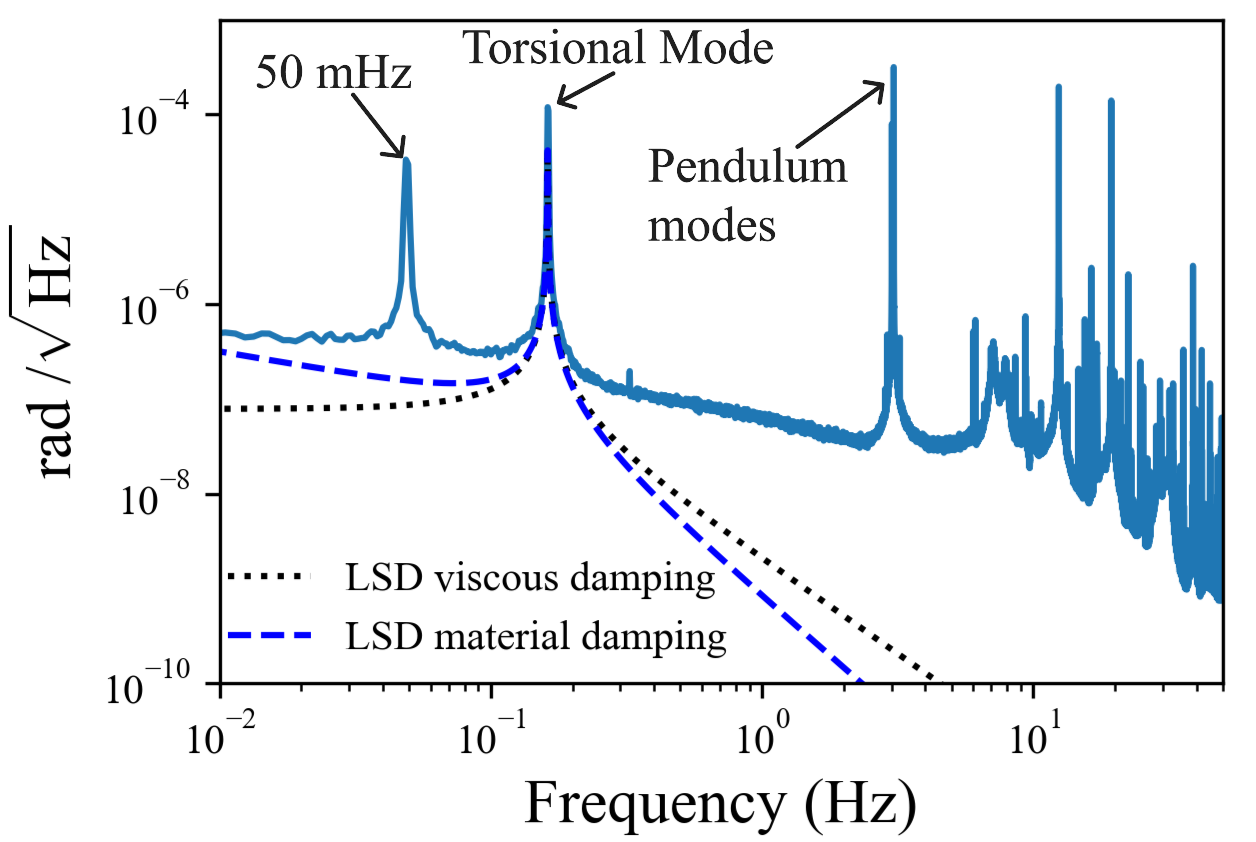}
        \caption{}
        \label{fig:PSDA}
    \end{subfigure}
    \hfill 
    \begin{subfigure}[]{0.50\textwidth} 
        \centering
        \includegraphics[width=\textwidth]{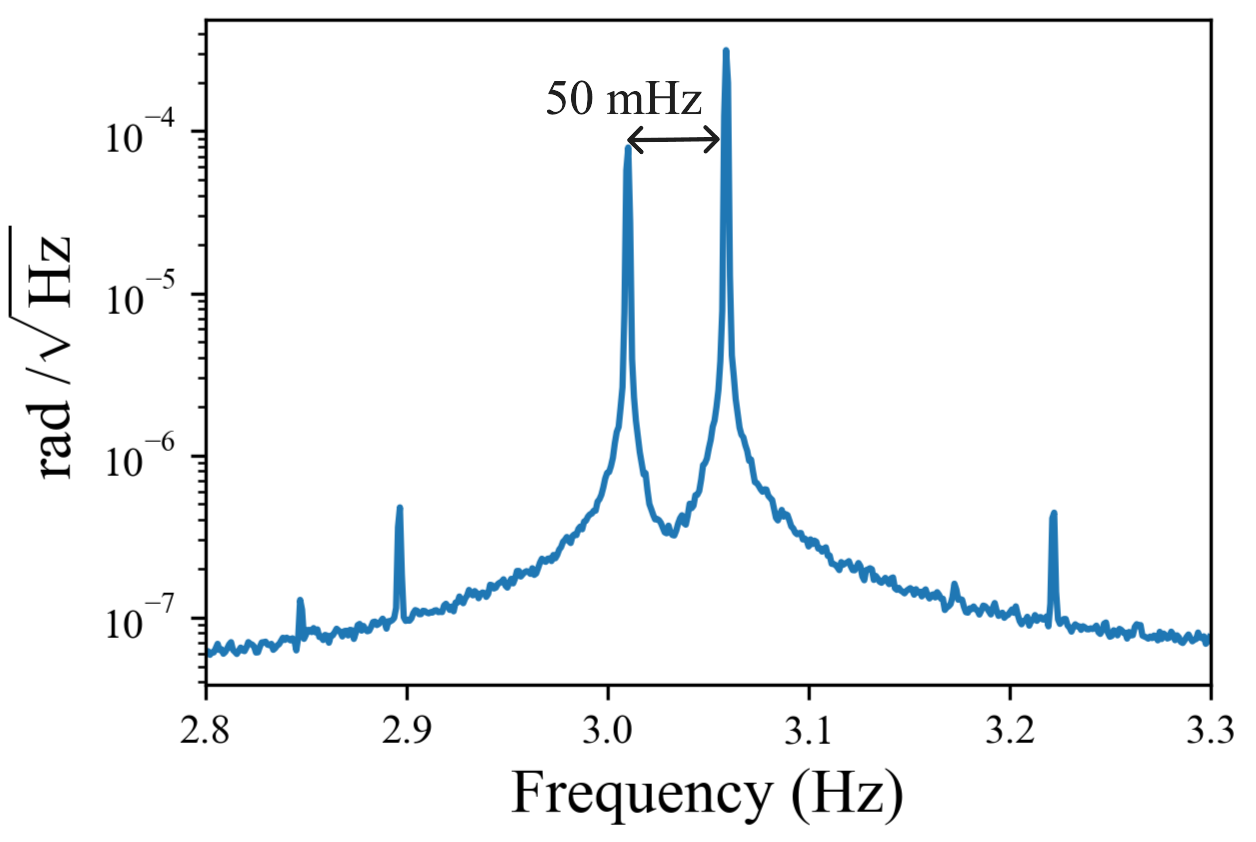}
        \caption{}
        \label{fig:PSDB}
    \end{subfigure}
    \caption{a) LSD of the displacement and thermal noise models. The dashed blue line is the thermal model assuming the damping is due to loss in the material, while the black dotted line is a thermal model assuming the damping is viscous damping.
    b) The same LSD zoomed into the pendulum modes.}
    \label{fig:PSD}
\end{figure*}

Figure \ref{fig:PSDA}, shows the linear spectral density (LSD) of the pendulum after it was allowed to ring down overnight. The black dotted curve is a thermal model of the pendulum experiencing gas damping only; the blue dashed curve is a thermal model where the damping is only through material loss\cite{saulson1990thermal}. Both thermal models were calculated at the natural frequency of the torsional pendulum.

Also apparent in Fig. \ref{fig:PSDA} is an oscillation below the torsional mode at 50 mHz. The 50 mHz frequency is attributed to a mixing of the two pendulum modes previously discussed: looking at the pendulum modes, Fig. \ref{fig:PSDB}, it is clear that the soft and stiff pendulum modes are present and their frequencies are separated by 50 mHz.

\subsection{\label{subsec:actuation} Optical Actuation}
Here we demonstrate the ability to optically actuate the torsional mode of the pendulum. The actuation is done with a delayed coordinate feedback control scheme built in LabVIEW\footnote{Certain equipment, instruments, software, or materials are identified in this paper in order to specify the experimental procedure adequately. Such identification is not intended to imply recommendation or endorsement of any product or service by the National Institute of Standards and Technology (NIST),nor is it intended to imply that the materials or equipment
identified are necessarily the best available for the purpose.}. The feedback input is the motion on the quadrant photodiode and output the same signal delayed by an amount which can be specified. The delayed signal is digitally filtered to single out the torsional mode frequency, which in turn is used to modulate actuation laser's power. The maximum laser power outside the vacuum chamber was measured to be approximate 0.6 mW. Accounting for the loss in power through the chamber viewport and the reflectivity of the silicon bob, and assuming normal incidence and a lever arm of 2.5 mm, the maximum actuating torque due to photon momentum transfer is approximated as 3$\times10^{-15}$ Nm. This torque would induce a static twist of 15 $\upmu$rad. We observed a larger displacement of about 37 $\upmu$rad in the signal during actuation, so photon momentum is not sufficient to completely account for the observation, suggesting that other affects may be present. See appendix \ref{append.A} for details.

\begin{figure*}
    \centering
    \begin{subfigure}[]{0.33\textwidth} 
        \centering
        \includegraphics[width=\textwidth]{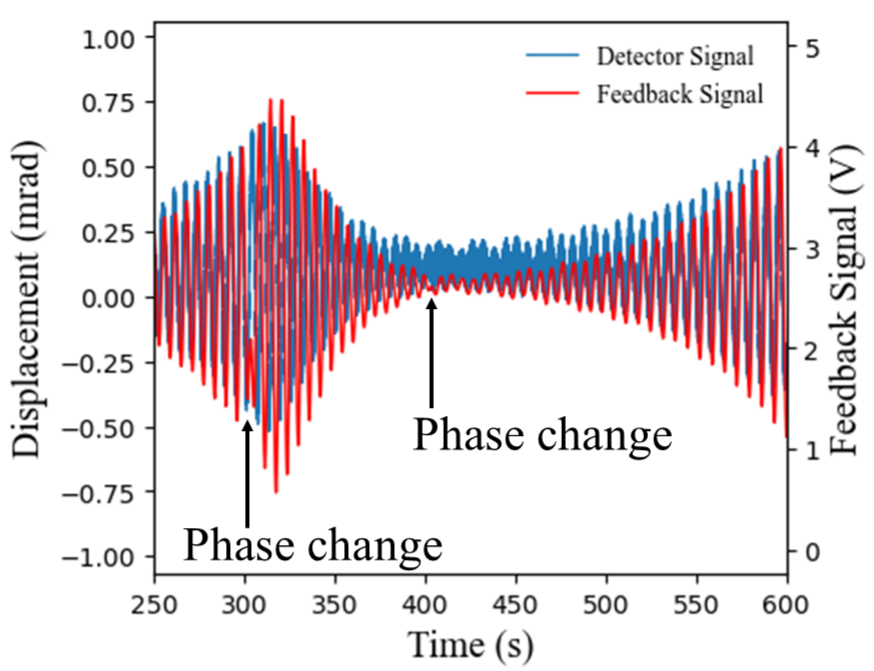}
        \caption{}
        \label{fig:actuation:phase}
    \end{subfigure}
    \hfill 
    \begin{subfigure}[]{0.33\textwidth} 
        \centering
        \includegraphics[width=\textwidth]{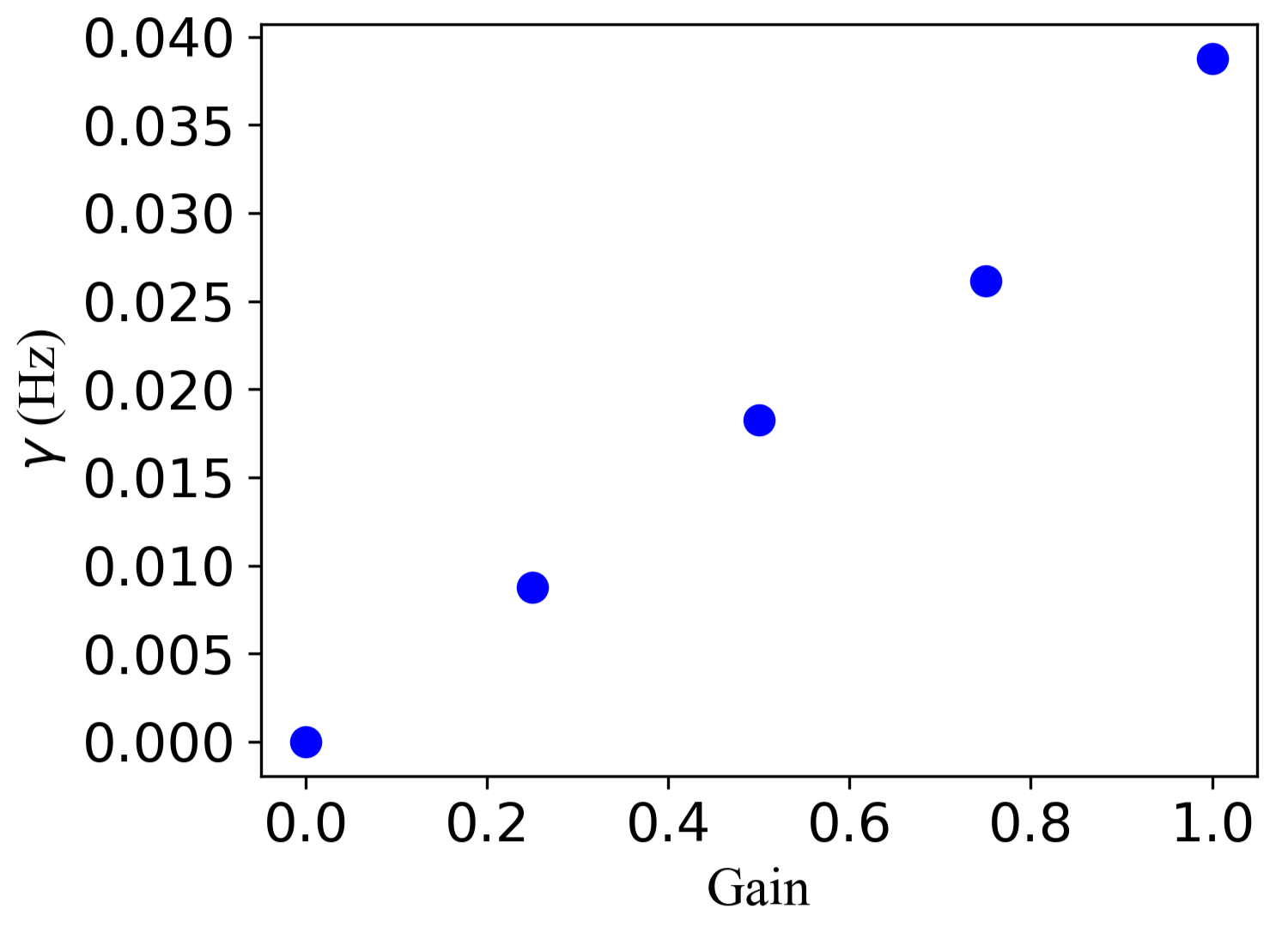}
        \caption{}
        \label{fig:gamma_gain_PSD_A}
    \end{subfigure}
    \hfill 
    \begin{subfigure}[]{0.33\textwidth} 
        \centering
        \includegraphics[width=\textwidth]{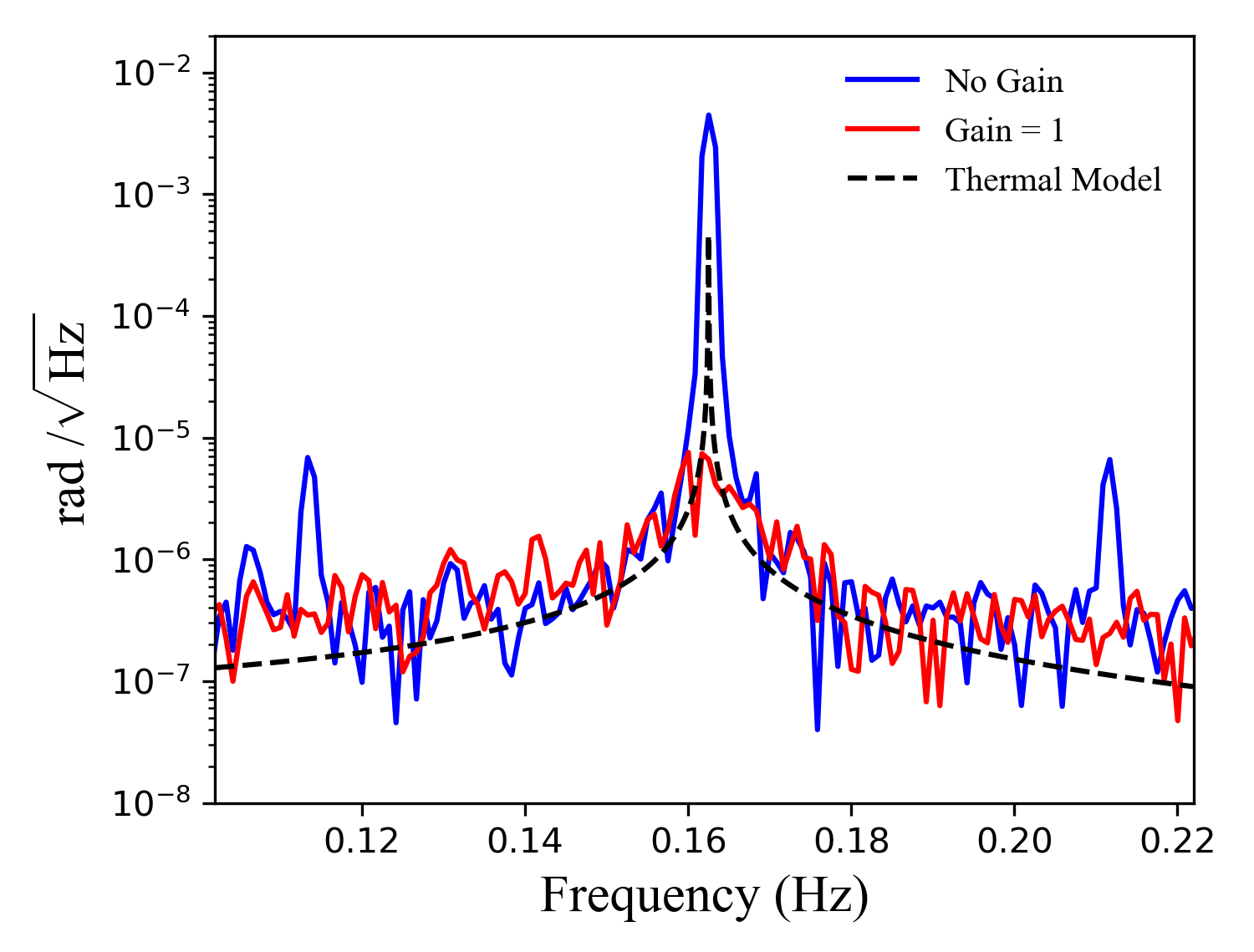}
        \caption{}
        \label{fig:gamma_gain_PSD_B}
    \end{subfigure}
    \caption{a) The feedback signal controls the damping or driving of the torsional mode by adjusting the phase between the detector signal and the feedback signal. 
    b) The damping rate as a function of the applied feedback gain. c) The LSDs of torsional mode for a free oscillation (blue) and an optically actuated oscillation with gain of 1 (red). The dashed black line is the thermal noise model. 
    }
\end{figure*}

The strength of the actuation is determined by the magnitude of the feedback gain. Whether the actuation excites or cools the pendulum is controlled by changing the delay and thus the phase between the readout signal and the actuation signal. Figure \ref{fig:actuation:phase} shows the detector signal, in blue, initially being driven by the feedback signal, in red. A phase change occurs at 300 s leading to the damping of the torsional mode, followed by another phase change to excite the torsional mode at 400 s. 

To show that the amount of damping can be controlled, the torsional mode of the oscillator was rung up to approximately the same amplitude 4 times and cooled over an hour at 4 different gain values. The damping rate was calculated for each of the gain values and, as seen in Fig. \ref{fig:gamma_gain_PSD_A}, the damping rate increases with increased gain. Furthermore, by cooling the pendulum with the maximum gain, we are able to suppress the torsional mode below the modeled thermal noise of the oscillator shown in Fig. \ref{fig:gamma_gain_PSD_B}. 

\section{\label{sec:} Conclusion}
This letter presents a unique single-end clamped Si$_3$N$_4$ torsional pendulum manufactured using wafer scale nanofabrication techniques that are scalable and capable of yielding multiple, identical copies. A low frequency and high intrinsic $Q$ were realized and measured in this first prototype. After it was successfully mounted in high vacuum, ring down tests were performed using a simple optical lever. We demonstrate and describe techniques and procedures to lithographically pattern torsion ribbons (25 mm long/0.0018 mm thick) that are continuous across a surrounding frame and a tethered mass. With practice and proper fixtures, the yield of pendulums during final release is near 100 \%. 

The combination of the observed torsional stiffness ($\kappa \approx 10^{-10}$ Nm/rad) and the low intrinsic loss of Si$_3$Ni$_4$ yields a high $Q$ torsional pendulum which is also
responsive to optical actuation using only the nitride membrane/silicon substrate interface as a mirror. Metal coatings or grating mirrors could be easily incorporated into the suspended paddle, improving our actuation abilities. 

Moving to a bifilar or high aspect ratio fiber configuration opens the possibility of increasing $Q$ due to the geometry. Likewise, mass loading of the pendulum can enable dissipation dilution due to gravity \cite{gonzfilez1994brownian,gonzalez2000suspensionsthermal}.
We point out that Si$_3$N$_4$ membranes are known to survive tensile stress well above 1 GPa \cite{ghadimi2018elastic}, so that the ribbon presented here could easily support a torsion beam mass on the order of 5 g, or over 100 times its present value. The fabrication procedure produces a monolithic pendulum up to the final attachment point that is readily amenable to well known bonding strategies, such as catalysis bonding\cite{elliffe2005hydroxide}, for the attachment of larger masses.

Our results suggest that the device-more broadly, mass-loaded Si$_3$N$_4$ nanosuspensions-hold promise for a challenging next generation weak force experiments.  In particular, optomechanical experiments in measurement-based feedback cooling, or gravitational experiments that seek to trap and manipulate a source or test mass in a potential that is on the order of two-body gravitational interactions.





\begin{acknowledgments}
This work was made possible by the Heising-Simons Foundation grant 2023-4467 “Testing the Quantum Coherence of Gravity”. The authors would also like to acknowledge Stephan Schlamminger for useful discussions and Robert Ilic and the NIST NanoFab for their expertise and help in fabrication. DJW and CAC acknowledge support from an RII UArizona National Labs Partnerships Grant. DJW acknowledges additional support from NSF through award no. 2239735.
\end{acknowledgments}
\newpage

\appendix

\section{Photon Momentum}
\label{append.A}
\begin{figure}[h]
    \begin{subfigure}{0.5\textwidth}
    \includegraphics[scale=0.25]{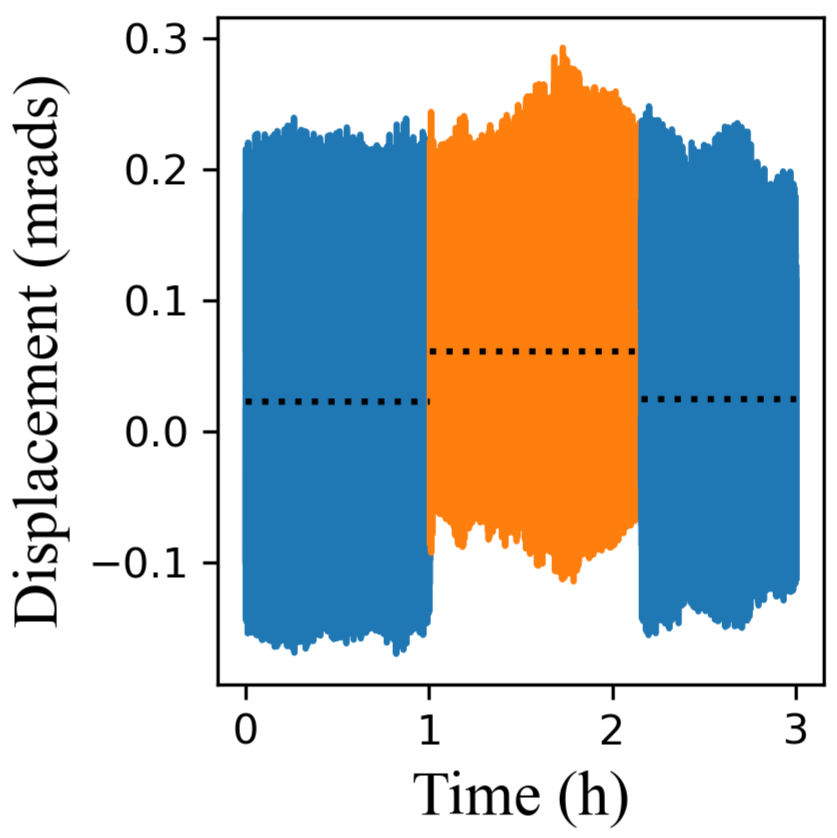}
    \end{subfigure}
    \caption{A static displacement of the pendulum is caused by the feedback laser and illustrated by the shift of the mean displacement signal. The feedback signal is on between the 1st and 2nd hour (the orange region) and off everywhere else. The dotted lines are the mean of the displacement within their respective regions.}
    \label{fig:offset}
\end{figure}

The maximum actuation torque due to photom momentum transfer was calculated using 
\begin{equation}
\tau_{photon} = \eta l\frac{2P_{max}}{c}\cos(\phi)
\label{eq:torque_photon}
\end{equation}
where $P_{max}$ is the maximum power of the actuating laser outside of the vacuum chamber, $l$ is the lever arm, $c$ is the speed of light, $\phi$ is the angle between bob's normal and the incident laser, and $\eta$ is a factor which accounts for loss of power through the chambers window and the silicon bob's reflectivity of 635 nm laser light; $\eta = 1$ would mean there is no power loss through the glass and the bob's surface was perfectly reflective. It is assumed that 96\% of the light transmits through the glass and that 35\% of that light is reflected by the silicon bob; thus for this work $\eta = 0.34$. Setting the torque from Eq. \ref{eq:torque_photon} equal to the restoring torque of the suspension, the displacement angle is calculated by
\begin{equation}
\theta = \frac{\tau_{photon}}{k_{tors}}
\end{equation}
where $\theta$ is the resulting displacement angle and $k_{tors}$ is the suspension stiffness given in the main text (Eq. \ref{eq.ktors}). The displacement resulting from the incident photons alone is calculated to be 15 $\upmu$rad, which is not enough to fully account for the deviation observed in Fig. \ref{fig:offset}. Taking the difference between the mean values recorded during actuation on and off yields a difference of 38 $\upmu$rad when going from actuation off to on and 36 $\upmu$rad when the actuation is turned off again.

Other factors, such as shifts in the center of gravity due to heating, or temperature induced strain in the ribbon likely contribute to the discrepancy, but could be easily remediated, for example with the deposition and patterning of a metal mirror on the paddle. 

\bibliography{citations}

\end{document}